\documentclass[aps,prl,twocolumn,groupedaddress,showpacs]{revtex4}
\usepackage{graphicx}% Include figure files
\usepackage{dcolumn}% Align table columns on decimal point
\usepackage{bm}% bold math

\newcommand{\bat}{$\bar{T}_w(i)$ }

\newcommand{\tallb}{$\bar{T}_{240}(i)$ }

\newcommand{\ovb}{$\Omega(w)$ }
\newcommand{\omax}{$\Omega_{max}$ }

\begin{document}

\title{Characteristic Length Scale of Electric Transport Properties of Genomes}

\author{C. T. Shih}
\affiliation{Department of Physics, Tunghai University, Taichung,
Taiwan}

\date{\today}

\begin{abstract}
A tight-binding model together with a novel statistical method are
used to investigate the relation between the sequence-dependent
electric transport properties and the sequences of protein-coding
regions of complete genomes. A correlation parameter $\Omega$ is
defined to analyze the relation. For some particular propagation
length $w_{max}$, the transport behaviors of the coding and
non-coding sequences are very different and the correlation
reaches its maximal value $\Omega_{max}$. $w_{max}$ and \omax are
characteristic values for each species. The possible reason of the
difference between the features of transport properties in the
coding and non-coding regions is the mechanism of DNA damage
repair processes together with the natural selection.
\end{abstract}

%\keywords{}

\maketitle

%\section{Introduction}
%\label{c:intro}

The conductance of DNA molecules is one of the central problems of
biophysics because it plays a critical role in the biological
systems. For example, it is postulated that there may be proteins
which can locate the DNA damage by detecting the long-range
electron migration properties\cite{rajski00,yavin05}. And for the
interest of applications, DNA is one of the most promising
candidates which may serve as the building block of molecular
electronics because of its sequence-dependent and self-assembly
properties.

There have been many experimental results on the conductance of
DNA from different measurements for the last few years. Yet the
results are still highly controversial\cite{endres04}. The
experimental results almost cover all possibilities, ranged from
insulating\cite{zhang02}, semiconducting\cite{porath00},
Ohmic\cite{tran00,yoo01}, and even induced
superconductivity\cite{kasumov01}. The diversity comes from the
methods of the measurements and the preparation of DNA samples.
One of the critical factors influencing the results is the contact
of the DNA and
electrodes\cite{zhang02,hartzell03,storm01,macia05}. The different
nucleotide sequences of the DNA molecules used in the experiments
also diversify the results because the transport properties are
sequence-dependent.

Aside from the electrical properties, the statistical features of
the symbolic sequences of DNA have also been studied intensely
during the past
years\cite{peng92,buldyrev95,li97,holste00,hsu03,messer05}. The
previous works are mainly focused on the correlations and
linguistic properties of the symbols A, T, C, and G, which
represent the four kinds of bases adenine, thymine, cytosine, and
guanine of the nucleotides, respectively. The analyses also give
some eccentric results. For example, the statistical behavior of
the intron-free coding sequences is similar to random sequences
while the intron-rich or junk sequences have long-range
correlations. One should note that the root of these statistical
properties of the symbolic sequences are the results of evolution,
and the underlying driving forces are the principles of physics
and chemistry. On the other direction, the correlation of
sequences will influence the physical and chemical properties,
such as the electric and mechanical properties of
DNA\cite{vaillant03}. Thus it is reasonable to conjecture that the
sequence-dependent electric properties can play critical roles
during the evolution process in nature by some ways such as the
DNA damage repair processes\cite{rajski00,yavin05}. In this
Letter, the relation between electric transport properties and the
gene-coding/nocoding parts of genomic sequences will be discussed.

The simplest effective tight-binding Hamiltonian for a hole
propagating in the DNA chain can be written
as\cite{roche03,roche03b}
\begin{equation}
H=\sum_n \epsilon_nc_n^{\dagger}c_n + \sum_n t_{n,n+1}(c_n^\dagger
c_{n+1} + H.C.) \label{e:model}
\end{equation}
where each lattice point represents a nucleotide base of the
chain. $c_n^\dagger$ ($c_n$) is the creation (destruction)
operator of a hole at the $n-$th site. $\epsilon_n$ is the
potential energy at the $n-$th site, which is determined by the
ionization potential of the corresponding nucleotide. $\epsilon_n$
equals to $8.24$ eV, $9.14$ eV, $8.87$ eV, and $7.75$ eV for $n=$
A, T, C, and G, respectively\cite{sugiyama96}. The DNA molecule is
assumed to be connected between two semi-infinite electrodes with
energy $\epsilon_m=\epsilon_G=7.75$ eV. The hopping integral
$t_{n,n+1}=t_m=1$ eV for electrodes and $t_{n,n+1}=t_{DNA}$ for
nucleotides. $t_{DNA}$ is assumed to be nucleotide-independent
here for simplicity. Typical value of $t_{DNA}=0.1\sim 0.4$ eV
from the first-principle calculation\cite{sugiyama96,zhang02b}. To
reduce the back scattering effect at the contacts, larger
$t_{DNA}$ (up to 1 eV) is also used in this study\cite{roche03}.
Note that $n\in(-\infty, 1]$ and $n\in[N+1,\infty)$ are for
electrodes and $n\in[2,N]$ are for nucleotides.

The eigenstates of the Hamiltonian $|\Psi\rangle = \sum_n a_n|
n\rangle$ ($|n\rangle$ represents the state that the hole is
located in the $n-$th site) can be solved exactly by using the
transfer matrix method:
\begin{eqnarray}
\left(
\begin{array}{c}
a_{N+2}\\ a_{N+1}
\end{array}
\right)=M_{N+1}M_N\cdots
M_1\left(\begin{array}{c}a_1\\a_0\end{array}\right)\equiv
P(N)\left(\begin{array}{c}a_1\\a_0\end{array}\right) \label{e:tm}
\end{eqnarray}
where
\begin{equation}
M_n=\left(\begin{array}{cc}\frac{E-\epsilon_n}{t_{n,n+1}} &
-\frac{t_{n-1,n}}{t_{{n,n+1}}}\\1 & 0\end{array} \right)
\end{equation}
$E$ is the energy of the injected hole. In electrodes, the wave
functions are plane waves and the dispersion of the hole is
$\epsilon_m+2t_mcosk$. So the range of possible $E$ is
$[\epsilon_m-2t_m, \epsilon_m+2t_m]=[5.75eV, 9.75eV]$. The
transmission coefficient has the following form\cite{macia99}
\begin{widetext}
\begin{equation}
T(E)=\frac{4-(\frac{E-\epsilon_m}{t_0})^2}{\sum_{i,j=1,2}P^2_{ij}+2-(\frac{E-\epsilon_m}{t_0})^2P_{11}P_{22}+(\frac{E-\epsilon_m}{t_0})(P_{11}-P_{22})(P_{12}-P_{21})}
\label{e:te}
\end{equation}
\end{widetext}

The transmission of several sequences of complete genomes
$S=(s_1,s_2,\cdots,s_{N_{tot}})$ is studied ($s_i=A$, $T$, $C$, or
$G$). Since the total length $N_{tot}$ of the complete genome is
usually much longer than the distance which holes can migrate
along the DNA chain even for the smallest $N_{tot}$ for viruses,
we won't measure the transmission through the whole chain but only
shorter segments instead. A ``window'' with width $w$ is defined
to extract a segment $S_{i,w}=(s_i,s_{i+1},\cdots,s_{i+w-1})$ for
$1\le i\le N_w=N_{tot}-w+1$ from $S$. Starting from $i=1$ and
sliding the window, we can get the ``transmission sequence''
$T_w(E,i)$ of $S_{i,w}$ for all $i$, which depends on the energy
of the injected hole $E$, the starting position of the segment
$i$, and the propagation length $w$. For further analysis of the
whole genome sequences, $T_w(E,i)$ is integrated in an energy
interval $[E,E+\Delta E$]:
\begin{equation}
\bar{T}_w(E,\Delta E,i)=\int_E^{E+\Delta E}T_w(E',i)dE'
\label{e:tm2}
\end{equation}
In the remaining of the Letter, the transmission is integrated for
the whole bandwidth, that is, $E=5.75$ eV and $\Delta E=4$ eV. And
these two values will be omitted in the related formulas for
short. 300 base pairs at the two ends of the DNA chain will be
omitted in the following analysis because the telomere sequences
at the terminals usually have larger transmission (due to the
periodicity) and will dominate some of the average properties.
Thus $N_w=N_{tot}-w+1-2\times 300$.

\begin{figure}[here]\rotatebox{-90}{\includegraphics[width=2.2in]{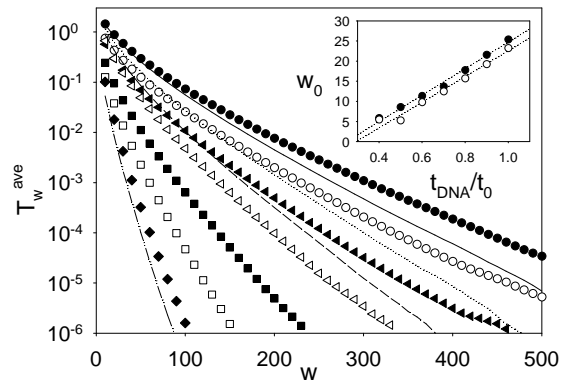}}
\caption{$T_w^{ave}$ for Y3 with $t_{DNA}=1.0$ eV (full circles),
$0.9$ eV (open circles), $0.8$ eV (full triangles), $0.7$ eV (open
triangles), $0.6$ eV (full squares), $0.5$ eV (open squares), and
$0.4$ eV (diamonds). Solid, dotted, dash, dash-dotted lines are
for a random sequence $R3$ with $t_{DNA}=1.0$, $0.9$, $0.8$ and
$0.4$ eV, respectively. (Inset) Localization length $w_0$ of Y3
(full circles) and R3 (open circles) for each $t_{DNA}$ (see
text).} \label{f:tw}
\end{figure}

The averaged transmission $T_w^{ave}=\frac{1}{N_w}\sum_i
\bar{T}_w(i)$ versus propagation length $w$ is plotted in
Fig.\ref{f:tw} for the third chromosome of Saccharomyces
cerevisiae (bakery yeast, accession number = NC.001135 for
GenBank\cite{benson04}, simplified as Y3 for short) with several
values of $t_{DNA}/t_0$. $T_w^{ave}$ decreases exponentially with
increasing $w$, which is consistent with the localization picture.
The curves can be fitted by the function $T_w^{ave}=ae^{-w/w_0}$.
The inset of Fig.\ref{f:tw} shows the averaged localization length
$w_0$ for each $t_{DNA}$. Note this is an averaged result of the
complete genome, and the possibility of high conductance of some
particular segments is not ruled out. Other important features are
that \bat decreases faster for smaller $t_{DNA}$, and $w_0$ is
nearly proportional to $t_{DNA}$. The reason is that the back
scattering is stronger for smaller $t_{DNA}$. Although smaller
$t_{DNA}$ ($\le 0.4$ eV) values are more physical, the signal
revealing the intrinsic properties of the sequences may be smeared
out by the strong back scattering. $T_w^{ave}$ for a random
sequence $R3$ with the same length and ratios of the four bases as
$Y3$ are also shown in the lines of Fig.\ref{f:tw}. It is clear
that the transmission of the random sequence decreases faster
(smaller $w_0$) than the natural genome due to the larger
disorder. This result is consistent with Ref.\cite{roche03}.

\begin{figure}[here]\rotatebox{-90}{\includegraphics[width=2.2in]{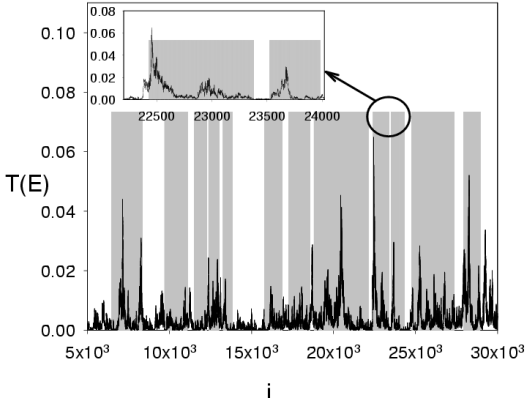}}
\caption{Comparison of \tallb (line, $t_{DNA}=1$ eV) and the
coding regions (shaded area) of the range from $5000-$th to
$30000-$th nucleotide of Y3. (Inset) Enlarged plot from $22000-$th
to $24000-$th nucleotide.} \label{f:y3_gene}
\end{figure}

Since the transport properties are related to the DNA damage
repair mechanism, there could be correlation between the locations
of genes and the corresponding integrated transmission \bat. In
Fig.\ref{f:y3_gene}, \tallb and the coding regions are compared
for part of the sequence of Y3. It seems that most of the sharp
peaks of \tallb are located in the protein-coding region.

To check this correlation in a more quantitative way, I first
define a binary ``coding sequence'' $G(i)=1$ $(0)$ if the $i-$th
nucleotide was in the protein-coding (noncoding) region, and then
normalize $G(i)$ and \bat in the following way
\begin{eqnarray}
G'(i)=G(i)-\frac{1}{N_w}\sum_j G(j);
g(i)=\frac{G'(i)}{\sqrt{\sum_j (G'(j))^2}}\nonumber
\end{eqnarray}
and
\begin{eqnarray}
\bar{T}'_w(i) = \bar{T}_w(i)-\frac{1}{N_w}\sum_j \bar{T}_w(i);
t_w(i) = \frac{\bar{T}'_w(i)}{\sqrt{\sum_j (\bar{T}'_w(i))^2}}
\label{e:normal}
\end{eqnarray}
The overlap between these two normalized sequences is defined
as\cite{shih00,shih02}
\begin{equation}
\Omega(w)=\sum_i g(i)t_w(i) \label{e:overlap}
\end{equation}

\begin{figure}[here]\rotatebox{-90}{\includegraphics[width=2.2in]{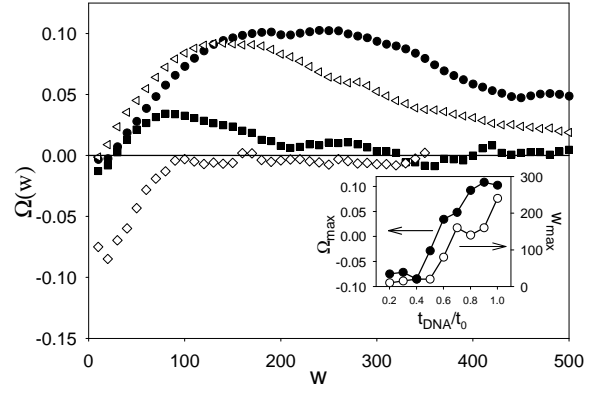}}
\caption{\ovb for $t_{DNA}/t_0=1.0$ (circles), $0.8$ (triangles),
$0.6$ (squares), and $0.4$ (diamonds) of Y3. (Inset) \omax (full
circles) and $w_{max}$ (open circles) as functions of $t_{DNA}$.}
\label{f:y3t}
\end{figure}

In Fig.\ref{f:y3t} \ovb for Y3 is shown for different $t_{DNA}$.
For $t_{DNA}=1$ eV, there is a maximum at $w_{max}=240$ with \omax
= $0.103$. Note that \omax denotes the maximal absolute value of
$\Omega(w)$ and can be positive or negative. The strong positive
overlap implies that the holes can move more freely in the coding
regions. As $t_{DNA}$ decreases, both \omax and $w_{max}$
decrease. For $t_{DNA}\le 0.5$ eV, the overlap becomes negative
which means the electronic conductance is poorer at the coding
regions. The dependence of \omax and $w_{max}$ on $t_{DNA}$ are
shown in the inset of Fig.\ref{f:y3t}. Although the values of
$w_{max}$ and \omax vary with $t_{DNA}$, $G(i)$ and \bat are
correlated in general.

\begin{figure}[here]\rotatebox{-90}{\includegraphics[width=2.2in]{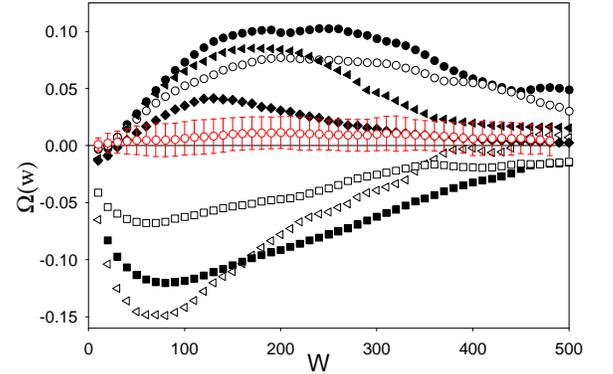}}
\caption{\ovb for several genomes: chromosomes III (full circles),
VIII (open circles), and X (full triangles) of yeast, Ureaplasma
parvum serovar 3 str. ATCC 700970 (full diamonds), acinetobacter
sp. ADP1 (full squares), Deinococcus radiodurans R1 chromosome II
(open triangles), and chlamydia trachomatis D/UW-3/CX (open
squares). Red circles with error bars are averaged $\Omega(w)$ for
10 randomized sequences of yeast chromosome III (see text).}
\label{f:over}
\end{figure}

Several \ovb with $t_{DNA}=1$ eV for different genomes are shown
in Fig.\ref{f:over}. It can be seen that there is maximal positive
or negative overlap $\Omega_{max}$ at some ``characteristic
migration length'' $w_{max}$ for each genome. \ovb for yeast
chromosomes III, VIII and X, and Ureaplasma parvum serovar 3 str.
ATCC 700970 are positive, which means the coding regions have
larger conductance. On the other hand, \ovb for acinetobacter sp.
ADP1, Deinococcus radiodurans R1 chromosome II, and chlamydia
trachomatis D/UW-3/CX are negative, which means the coding regions
have smaller conductance. $(\Omega_{max},w_{max})$ for these
genomes are summarized in TABLE.\ref{t:omax}.

\begin{table}
\caption{\label{t:omax}\omax and $w_{max}$ for the genomes studied
in Fig.\ref{f:over}.}
\begin{ruledtabular}
\begin{tabular}{lccr}
Genome & Access No. & $w_{max}$ & $\Omega_{max}$\\ \hline Yeast
III & NC.001135 & 240 & 0.103 \\ Yeast VIII & NC.001140 & 200 &
0.077\\ Yeast X & NC.001142 & 170 & 0.085 \\Ureaplasma parvum
 & NC.002162 & 130 & 0.041 \\serovar 3 str. ATCC 700970 \\ \hline acinetobacter sp.
 ADP1 & NC.005966 & 80 & -0.129 \\ Deinococcus radiodurans &
 NC.001264 & 80 & -0.149\\ R1
chromosome II\\chlamydia trachomatis & NC.000117 & 50 &
-0.075\\D/UW-3/CX\\
\end{tabular}
\end{ruledtabular}
\end{table}

To ensure that \ovb shown above are physically and biologically
meaningful, we compare the results with random sequences. Ten
sequences generated by the same way as $R3$ are analyzed and the
averaged \ovb (overlap with the $g(i)$ of Y3) are shown in
Fig.\ref{f:over} (open circles with error bars). It is clear that
its overlap is about one order of magnitude smaller then the real
sequences. So $\Omega_{max}$ and $w_{max}$ are not artifacts, but
intrinsic properties of genomes from the above comparison.

From the analysis above, it can be concluded that $w_{max}$ is a
characteristic length scale of the electric transport, which can
make out the gene-coding regions. And \omax stands for the
``sensibility'' of this probing process.

The possible biological reason of these correlations is the
mechanism of DNA damage repair processes. Since proteins use the
transport properties to probe the location of DNA
damage\cite{rajski00,yavin05}, the transport of the coding areas
should have particular features for the detecting processes, while
those of the non-coding regions are somewhat irrelevant.

Fig.\ref{f:over} shows two important features of $\Omega_{max}$.
First, each species has their characteristic values
$(w_{max},\Omega_{max})$. It can be postulated that the mechanisms
detecting the defects of DNA of different species are different
due to the various biological and environmental features. Second,
$(w_{max},\Omega_{max})$ of the different chromosomes of the same
species (yeast here) are very similar because they are in the {\it
same} environment, hence the same DNA damage repair mechanism.

It should be noted that the model used in this study is an
oversimplified one. However, one of the most important properties
can be extracted from this coarse-grained model -- the coding
regions have very different transport behavior from the noncoding
parts at the characteristic length scale $w_{max}$. And each
species has different $w_{max}$ to adjust their environment. In
the future, the model will be finer-grained by introducing the
more realistic interactions like the base-dependent
hopping\cite{klotsa05}, the sequence dependent
potentials\cite{senthilkumar03}, and the charge-charge
interactions.

In summary, with a new method combining the transfer matrix
approach and symbolic sequence analysis, the correlation between
the transport properties and the positions of genes is studied for
complete genomes. There are two characteristic values
$\Omega_{max}$ and $w_{max}$ for each genome. These two values can
provide information for taxonomy or the mechanism of evolution.

This research is supported by the National Science Council in
Taiwan (Grant No.93-2112-M-029-001-). Part of the calculations are
performed in the National Center for High-performance Computing in
Taiwan. The author is grateful for their help.

\end{document}